\documentclass[12pt]{article}
\usepackage{amssymb,amsmath,epsfig}
\allowdisplaybreaks

\begin{document}
\title{\bf Greybody Factor for a Rotating Bardeen Black Hole by Perfect Fluid Dark Matter}
\author{M. Sharif \thanks{msharif.math@pu.edu.pk} and Sulaman Shaukat\thanks{sulamanshaukat444@gmail.com}\\
Department of Mathematics, University of the Punjab,\\
Quaid-e-Azam Campus, Lahore-54590, Pakistan.}

\date{}
\maketitle
\begin{abstract}
In this paper, the greybody factor is studied analytically for a
rotating regular Bardeen black hole surrounded by perfect fluid dark
matter. Firstly, we examine the behavior of effective potential by
using the radial equation of motion developed from the Klein-Gordon
equation. We then consider tortoise coordinate to convert the radial
equation into Schr\"odinger form equation. We solve the radial
equation of motion and obtain two different asymptotic solutions in
terms of hypergeometric function measured at distinct regimes so
called near and far-field horizons. These solutions are smoothly
matched over the whole radial coordinate in an intermediate regime
to check their viability. Finally, we measure the absorption
probability for massless scalar field and examine the effect of
perfect fluid dark matter. It is concluded that both the effective
potential and greybody factor increase with perfect fluid dark
matter.
\end{abstract}

\section{Introduction}

Black holes (BHs) are very interesting astronomical compact objects
with a strong gravitational pull that nothing can escape from it not
even light. These dense objects contain event horizon as well as
singularity. The theory of general relativity (GR) describes
singularity-free BH solutions which are asymptotically flat as well
as static spherically symmetric spacetimes named as regular BHs.
Bardeen \cite{1} was the first to calculate the singularity-free
solution of spherically symmetric BH referred to as Bardeen regular
BH. Hayward \cite{2} extended the Bardeen concept of regular BHs to
develop a non-rotating regular BH. Later, Bambi and Modesto \cite{3}
applied the Newman-Janis procedure to Bardeen  as well as Hayward
BHs and introduced the family of rotating regular BH solutions.

Our universe contains a bulk amount of non-radiating matter
distribution named as dark matter which is composed of unfamiliar
subatomic particles. The word dark indicates that it does not
reflect, absorb or emit electromagnetic radiation and cannot be
observed directly. Dark matter contents which preserve the
properties of perfect fluid such as an isotropic pressure and mass
density are called perfect fluid dark matter (PFDM). It is
worthwhile to construct the solutions of BH surrounded by dark
matter as well as dark energy. The study of BHs in the presence of
quintessence type dark energy has become an interesting topic in the
last two decades. Kiselev \cite{4} was the first to propose the
uncharged and charged solutions of Schwarzschild BH surrounded by
quintessence matter. There are different types of BH solutions
constructed for quintessence field by taking into account of
Kiselev's idea \cite{5}. Xu et al. \cite{6} used Newman-Janis
algorithm on Kerr-like BH to find spherically symmetric BH solution
in the presence of PFDM. They also extended the solution for the
Kerr-de Sitter/anti-de Sitter spacetimes with cosmological constant.
Xu et al. \cite{7} generalized Reissner-Nordstr\"om spacetime to the
Kerr-Newman-anti-de Sitter spacetime with PFDM. They also
investigated that the BH singularity does not change with the
effects of PFDM. Hou et al. \cite{8} studied the influence of PFDM
to examine phase transition as well as thermodynamics for the
Reissner-Nordstr\"om-anti-de Sitter BH.

Hawking \cite{9} found that thermal radiations are generated and
then released from BHs due to quantum mechanical effects known as
Hawking radiations. These radiations gradually decrease the mass of
BHs and eventually contribute to its evaporation, i.e., any BH
having mass less than $10^{15}$g would vanish. In terms of
frequency, the emission rate of BH at the event horizon can be
calculated as
\begin{equation}\nonumber
\gamma(w)=\left(\frac{d^{3}\kappa}{8\pi^3e^{\frac{w}{T_{H}}}}\right),
\end{equation}
where $\kappa$, $w$ and $T_{H}$ denote the surface gravity, wave
frequency and Hawking temperature, respectively. It is equally valid
for both massive as well as massless particles and can also be used
for rotating/non-rotating BHs. The emission rate of particles is
significantly affected by the event horizon because it behaves as a
barrier to filter the Hawking radiations. The spectrum of radiations
at event horizon is similar to black body while the radiation
spectra is different observed by a distant observer. There is a
non-trivial spacetime around a BH that changes the Hawking radiation
spectra as some radiations reflect back to BH and rest of them cross
the barrier. The probability of absorption rate of waves coming from
infinity and absorbed by the BH (proportional to the area of
absorption cross-section) is called greybody factor
\cite{13}-\cite{16}. The following expression is a relation between
greybody factor and emission rate of BH as
\begin{equation}\nonumber
\gamma(w)=\left(\frac{ d^{3}\kappa |{\emph{A}}_{l,m}|^{2}}{8\pi^3
e^{\frac{w}{T_{H}}}}\right),
\end{equation}
here $|{\emph{A}}_{l,m}|^{2}$ is called the greybody factor that
depends on the frequency of massless particles.

Creek et al. \cite{17} calculated the greybody factor for rotating
BHs to investigate the emission rate of the scalar field through
analytical as well as numerical solutions. Some rigorous limits to
greybody factor are determined by Boonserm et al. \cite{18} for
Myers-Perry BHs. Jorge et al. \cite{19} computed the greybody factor
for higher-dimensional rotating BHs with the cosmological constant
in a low-frequency regime. Toshmatov et al. \cite{20} measured the
effect of charge and absorption rate for regular BHs. It is found
that the presence of charge reduces the transmission factor for
incident waves. Ahmed and Saifullah \cite{21} used cylindrical
symmetric spacetime to find the analytic solution of greybody factor
for uncharge massless scalar field. Hyun et al. \cite{22} applied
the spheroidal joining factor for rotating BH and found the analytic
solution of the greybody factor using brane scalar fields. Dey and
Chakrabarti \cite{23} considered Bardeen-de Sitter spacetime and
measured the absorption probability as well as quasinormal modes.

Ida et al. \cite{24} examined the greybody factor using brane scalar
field for rotating BH in a low-frequency expansion. Chen et al.
\cite{25} calculated the greybody factor for d-dimensional BH using
quintessence field and found that by increasing $|w_{q_{\ast}}|$,
the luminosity of radiation decreases monotonically. They also
explained that the corresponding solution reduces to the
d-dimensional Reissner-Nordstr\"om BH for
$w_{q_{\ast}}=\frac{d-3}{d-1}$. Crispino et al. \cite{26}
investigated the greybody factor and absorption process of
Schwarzschild BH for non-minimally coupled scalar fields. Kanti et
al. \cite{27} derived the greybody factor for scalar field using
higher-dimensional Schwarzschild-de Sitter spacetime in a low energy
regime. Ahmed and Saifullah \cite{28} illustrated the greybody
factor for charged BH in the presence of cosmological constant.
Sharif and Ama-Tul-Mughani \cite{29} examined the greybody factor
for rotating Bardeen BH and Kerr-Newman BH surrounded by
quintessence.

Sakalli \cite{36a} investigated the problems of resonant
frequencies, entropy/area quantization, and greybody factor of the
rotating linear dilaton BH. Sakalli and Aslan \cite{37a} computed
the exact greybody factor, the absorption cross-section and the
decay rate for the massless scalar waves of non-asymptotically flat
rotating linear dilaton BHs. Kanzi and Sakalli \cite{39a} studied
the effect of Lorentz symmetry breaking on the Hawking radiation and
computed the semi-analytic greybody factor (for both bosons and
fermions)of Schwarzschild-like BH found in the bumblebee gravity
model. Gursel and Sakalli \cite{40a} studied the greybody factor,
absorption cross-section, and decay rate of the non-Abelian charged
Lifshitz black branes. Recently, Jusufi et al \cite{41a} studied
quasinormal modes in 5D electrically charged Bardeen BHs spacetime
by considering the scalar and electromagnetic field perturbations.
They showed that the transmission (reflection) coefficients decrease
(increase) with an increase in the magnitude of the electric charge.

There is a large body of literature to study the effects of PFDM on
the geometry as well as physical characteristics of BH spacetimes.
Rahaman et al. \cite{30} used dark matter as perfect fluid in the
flat rotation curve and found its general features through the
equation of state of PFDM. Hou et al. \cite{31} observed the shadow
of Saggitarius $A^{\star}$ located at the mid of our Milky Way and
also examined the effects of rotation and PFDM parameters. They also
investigated the emission rate of energy for different values of the
PFDM parameter. Jamil et al. \cite{32} calculated the solutions of
rotating as well as non-rotating BHs in the presence of PFDM and
discussed their null geodesics. Hendi et al. \cite{33} examined the
rotating BH surrounded by PFDM and investigated the phase
transitions as well as its instability. Das et al. \cite{34} found
the solution of charged BH in the presence of PFDM and also studied
its circular geodesics. Recently, Ama-Tul-Mughani et al. \cite{35}
investigated the greybody factor as well as effects of thermal
fluctuations on thermodynamics of non-rotating regular Bardeen BH
surrounded by PFDM.

In this paper, we explore the effective potential and greybody
factor for rotating regular Bardeen BH surrounded by PFDM. The paper
is organized as follows. Section \textbf{2} contains the formation
of effective potential by using the radial equation of motion. We
study the analytic solution of the greybody factor for two distinct
regimes by solving the radial equation in section \textbf{3}. In
section \textbf{4}, we compare both the solutions and finally
compute the absorption as well as emission rates for the massless
scalar field. In the last section, we summarize the obtained
results.

\section{Effective Potential}

The spacetime for rotating Bardeen BH with PFDM is given as
\cite{36}
\begin{equation}\label{1}
ds^{2}=-E(r,\theta)dt^{2}+F(r,\theta)dr^{2}+G(r,\theta)d\theta^{2}+I(r,\theta)d\phi^{2}-4J(r,\theta)dtd\phi,
\end{equation}
where
\begin{eqnarray}\nonumber
E(r,\theta)&=&1-\frac{2\rho r}{G(r,\theta)},\quad
F(r,\theta)=\frac{G(r,\theta)}{\Omega(r)}, \quad
G(r,\theta)=a^{2}\cos^{2}\theta+r^{2},\\\nonumber
I(r,\theta)&=&\sin^{2}\theta\bigg[r^{2}+a^{2}+\frac{2\rho
r\sin^{2}\theta a^{2}}{G(r,\theta)}\bigg],\quad
J(r,\theta)=\frac{a\rho r\sin^{2}\theta}{G(r,\theta)},\\\nonumber
2\rho&=&\frac{2Mr^{3}}{(r^{2}+q^{2})^{\frac{3}{2}}}-\alpha\ln\frac{r}{|\alpha|},\quad
\Omega(r)=r^{2}+a^{2}-\frac{2Mr^{4}}{(r^{2}+q^{2})^{\frac{3}{2}}}+\alpha
r\ln\frac{r}{|\alpha|}.
\end{eqnarray}
Here $\alpha$ is a PFDM parameter related to density and pressure
and $a$, $M$, $q$ denote the rotation parameter, gravitational mass,
magnetic charge of BH, respectively. The line element (\ref{1})
reduces to the Schwarzschild spacetime if  $a=0$, $\alpha=0$ and
$q=0$. We can obtain the inner ($r_{-}$) and outer ($r_{+}$) event
horizons by taking $\Omega(r)=0$, i.e.,
\begin{equation}\label{2}
a^{2}+r^{2}-\frac{2Mr^{4}}{(r^{2}+q^{2})^{\frac{3}{2}}}+\alpha
r\ln\frac{r}{|\alpha|}=0.
\end{equation}

Now we determine the greybody factor analytically. For this purpose,
we first calculate the equation of motion to analyze the propagation
of massless scalar field. It is assumed that particles are minimally
coupled to gravity and they do not have any other type of
interaction. In this background, the equation of motion turns out to
be
\begin{equation}\label{3}
\nabla_{\beta}\nabla^{\beta}\Psi=\partial_{\beta}[\sqrt{-g}g^{\beta\gamma}\partial_{\gamma}\Psi]=0,
\end{equation}
where $\Psi=\Psi(t,r,\theta,\phi)$ is a massless scalar field.
Inserting the values from Eq.(\ref{1}), it follows that
\begin{eqnarray}\nonumber
&&\sqrt{-g}\bigg(\frac{-I}{IE+4J^{2}}\bigg)\partial_{tt}\Psi-4\sqrt{-g}
\bigg(\frac{-K}{IE+4J^{2}}\bigg)\partial_{t}\partial_{\phi}\Psi+
\bigg(\frac{1}{F}\sqrt{-g}\partial_{r}\Psi\bigg)_{,r}
\\\label{4}&&+\bigg(\frac{1}{G}\sqrt{-g}\partial_{\theta}\Psi\bigg)_{,\theta}+
\sqrt{-g}\bigg(\frac{F}{IE+4J^{2}}\bigg)\partial_{\phi\phi}\Psi =0.
\end{eqnarray}
We note that
\begin{equation}\nonumber
IE+4J^{2}=\Omega(r)\sin^{2}\theta.
\end{equation}
Using the separation of variables method, we can write
\begin{equation}\nonumber
\Psi=\exp(\iota m\phi)\exp(-\iota w t)R_{w l m}(r)T^{m}_{l}(\theta,a
w),
\end{equation}
where $T^{m}_{l}(\theta,a w)$ is the angular spheroidal function.
Thus Eq.(\ref{4}) can be written into radial and angular equations
of motion as \cite{37}
\begin{eqnarray}\nonumber
&&\frac{\partial }{\partial r}\bigg[\Omega\frac{\partial R_{w l
m}}{\partial
r}\bigg]+\bigg[\frac{1}{\Omega}\bigg(w^{2}(a^{2}+r^{2})^{2}+a^{2}
m^{2}-2 a m w(a^{2}+r^{2}-\Omega)\bigg)\\\label{5}&&-a^{2} w^{2}-
\lambda^{m}_{l}\bigg]R_{w l m}(r)=0,
\end{eqnarray}
and
\begin{eqnarray}\nonumber
&&\frac{1}{\sin\theta}\bigg[\frac{\partial}
{\partial\theta}\bigg(\sin\theta\frac{\partial
T^{m}_{l}}{\partial\theta}\bigg)-\bigg(-
a^{2}w^{2}\sin\theta\cos^{2}\theta+\frac{m^{2}}
{\sin\theta}-\lambda^{m}_{l}\sin\theta\bigg)\nonumber\\&&\times
T^{m}_{l}(\theta,a w)\bigg]=0\nonumber,
\end{eqnarray}
respectively. Here $\lambda_{l}^{m}$ corresponds to the separation
constant that explains a relationship between the decoupled
equations.

In general, the separation constant cannot expressed in a closed
form. However, we can write its analytical solution in series form
with parameter $aw$ given as \cite{38}
\begin{equation}\nonumber
\lambda_{l}^{m}=\sum_{n=0}^{\infty}(aw)^{n}\textit{f}_{~n}^{~lm}.
\end{equation}
For the sake of convenience, we break up the series and only retain
upto the third order terms  
\begin{equation}\nonumber
\lambda^{m}_{l}=(l^{2}+l)+\frac{2m^{2}-2(l^{2}+l)+1}{(2l+3)(2l-1)}
(a w)^{2}+O((aw)^{4}),
\end{equation}
as $\textit{f}_{~1}^{~lm}=\textit{f}_{~3}^{~lm}=0$. Here $l$ is the
orbital angular momentum satisfying the relation $|m|\leqslant l$
with non-negative values. Using this power series expansion, we can
solve radial equation of motion (\ref{5}) analytically. The
resulting solution yields the greybody factor for a massless scalar
field. First, we investigate the profile of effective potential
(responsible for the greybody factor) by solving the above radial
equation. We define a new radial transformation as
\begin{equation}\nonumber
R_{wlm}(r)=\frac{S_{wlm}(r)}{\sqrt{a^{2}+r^2}}.
\end{equation}
Using tortoise coordinate $t_{*}$, we have
\begin{equation}\nonumber
\frac{dt_{*}}{dr}=\frac{a^{2}+r^2}{\Omega(r)},
\end{equation}
such that
\begin{equation}\nonumber
\frac{d}{dt_{*}}=\frac{\Omega(r)}{a^{2}+r^2}\frac{d}{dr},\quad
\frac{d^2}{dt_{*}^{2}}=\left(\frac{\Omega(r)}{a^{2}+r^2}
\right)\bigg(\left(\frac{\Omega(r)}{a^{2}+r^2}\right)
\frac{d^2}{dr^2}+\frac{d}{dr}
\left(\frac{\Omega(r)}{a^{2}+r^2}\right)\frac{d}{dr}\bigg).
\end{equation}
We note that as $r$ approaches to $r_{h}$, $t_{*}\rightarrow
-\infty$ and for $r\rightarrow\infty$, $t_{*}\rightarrow \infty$.
Therefore, the model changes its range from $-\infty$ to $+\infty$
due to the tortoise coordinate $t_{*}$ whereas Eq.(\ref{5}) is
limited to the regions located outside the BH horizon. We can write
Eq.(\ref{5}) in the form of Schr\"odinger wave equation as
\begin{equation}\nonumber
(\frac{d^2}{dt_{*}^2}-V_{eff})S_{wlm}(r)=0,
\end{equation}
where the effective potential is given by
\begin{eqnarray}\nonumber
V_{eff}&=&\bigg[\sqrt{a^{2}+r^2}\frac{d}{d r}\bigg(\frac{r
\Omega(r)}{(a^{2}+r^2)^{\frac{3}{2}}}\bigg)-\frac{1}{\Omega}
\bigg(w^{2}(a^{2}+r^2)^{2}+a^{2}
m^{2}\\\nonumber&-&2amw(a^{2}+r^2-\Omega)\bigg)+ a^{2}
w^{2}+(l^{2}+l)\\\nonumber&+&
\frac{2m^{2}-2(l^{2}+l)+1}{(2l+3)(2l-1)}(a
w)^{2}\bigg]\frac{\Omega(r)}{(a^{2}+r^2)^{2}}.
\end{eqnarray}

When $\Omega(r)=0$, the effective potential vanishes at the event
horizon. The effective potential works as a potential barrier and
its graphical analysis against $\frac{r}{r_{h}}$ is analyzed for
different values of physical parameters in Figures \textbf{1-3}.
Initially, we consider $\alpha=0.01$, $M=m=l=1$, $w=0.1$ and draw
graphs for various values of the rotation parameter. The left and
right plots of Figure \textbf{1} show the effects of magnetic charge
and rotation parameter on the potential barrier, respectively. It is
found that the barrier's height increases by increasing the value of
$q$ and ultimately enhances the absorption rate (left plot). The
barrier height decreases by increasing the values of $a$ which
reduces the emission rate of the scalar field (right plot). This
shows that the greybody factor significantly increases for large
values of $a$. Figure \textbf{2} represents the effect of angular
momentum on the potential barrier. This indicates that the effective
potential grows rapidly for higher values showing the failure of
emission of massless scalar field particles. We also investigate the
effect of PFDM parameter over the potential function graphically.
Figure \textbf{3} shows that the effective potential increases for
positive values of PFDM parameter while it decreases for negative
values.
\begin{figure}\center
\epsfig{file=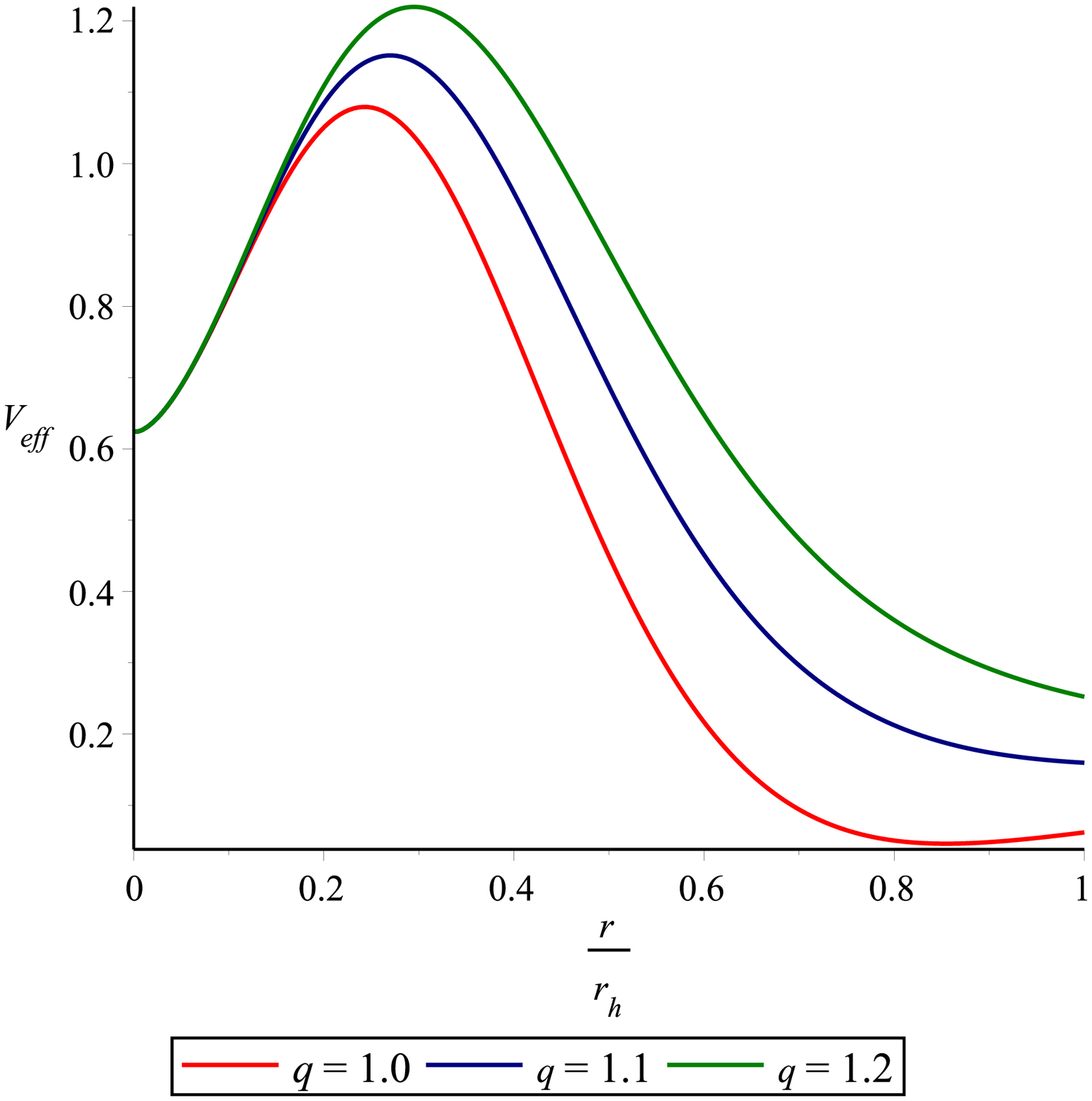,width=0.5\linewidth}\epsfig{file=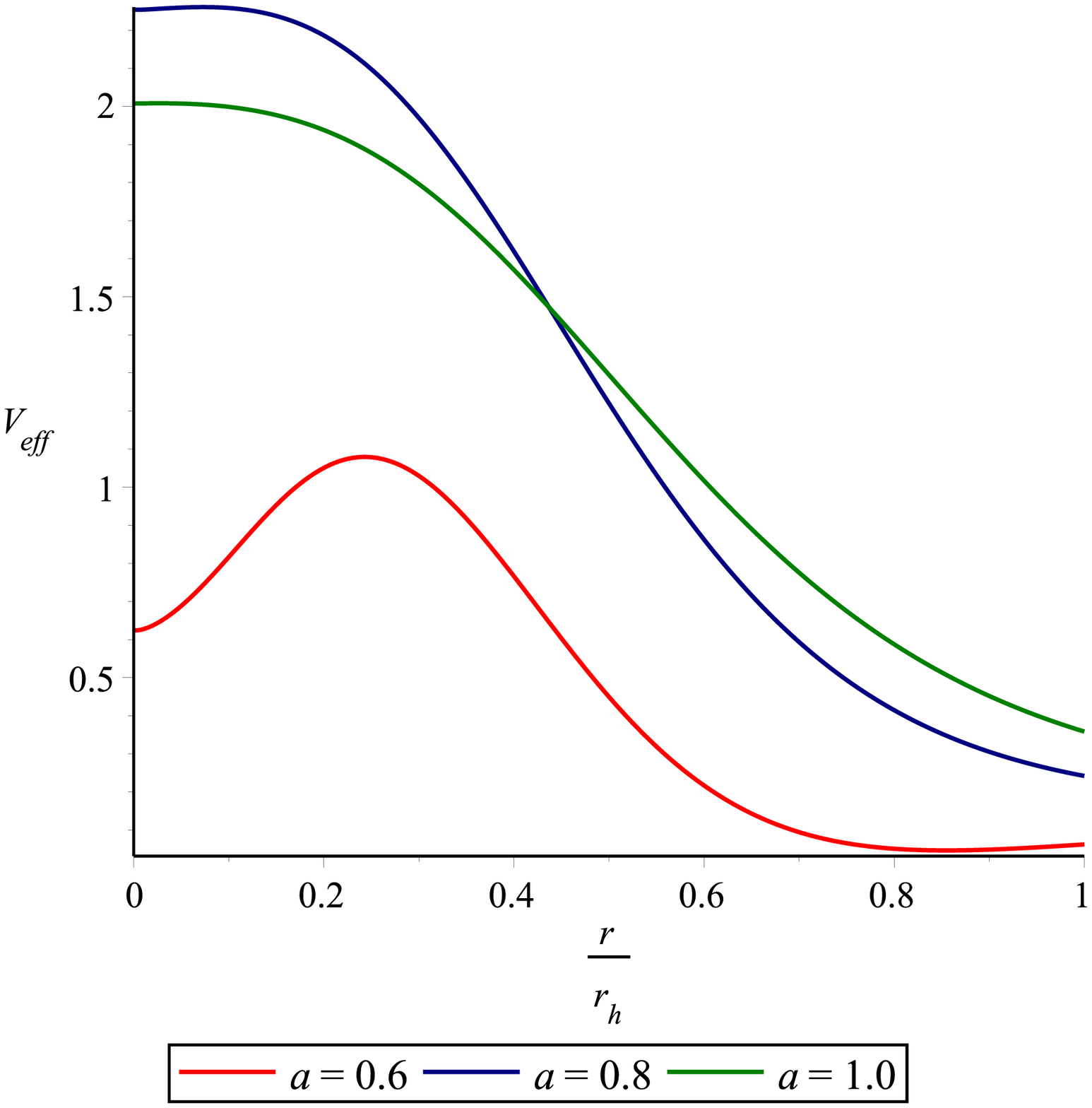,width=0.5\linewidth}
\caption{Effective potential for massless scalar field for $a=0.6$
(left) and $q=1.0$ (right) with $m=1$, $M=1$, $l=1$, $\alpha=0.01$
and $w=0.1$.}
\end{figure}
\begin{figure}\center
\epsfig{file=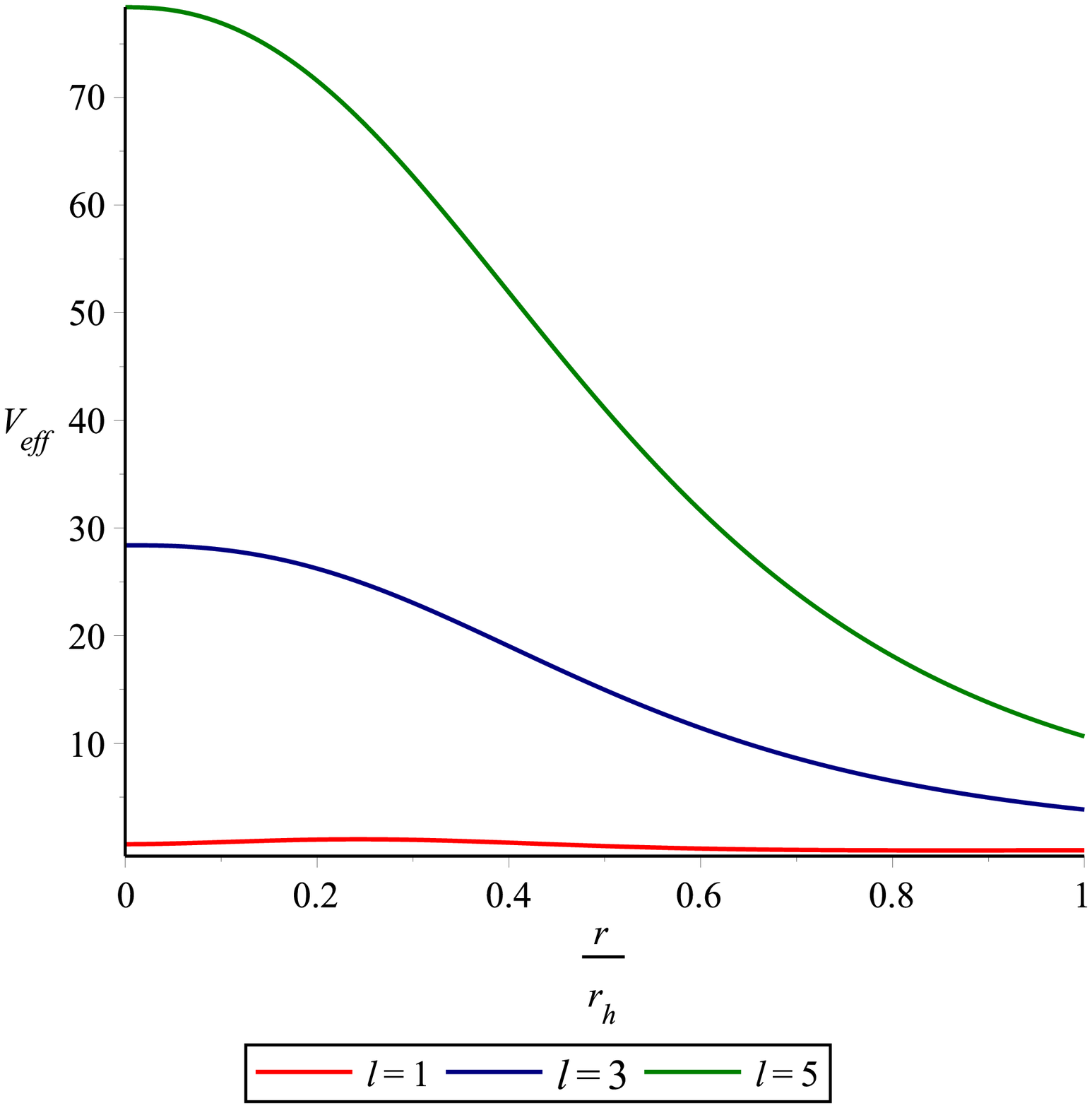,width=0.5\linewidth}\epsfig{file=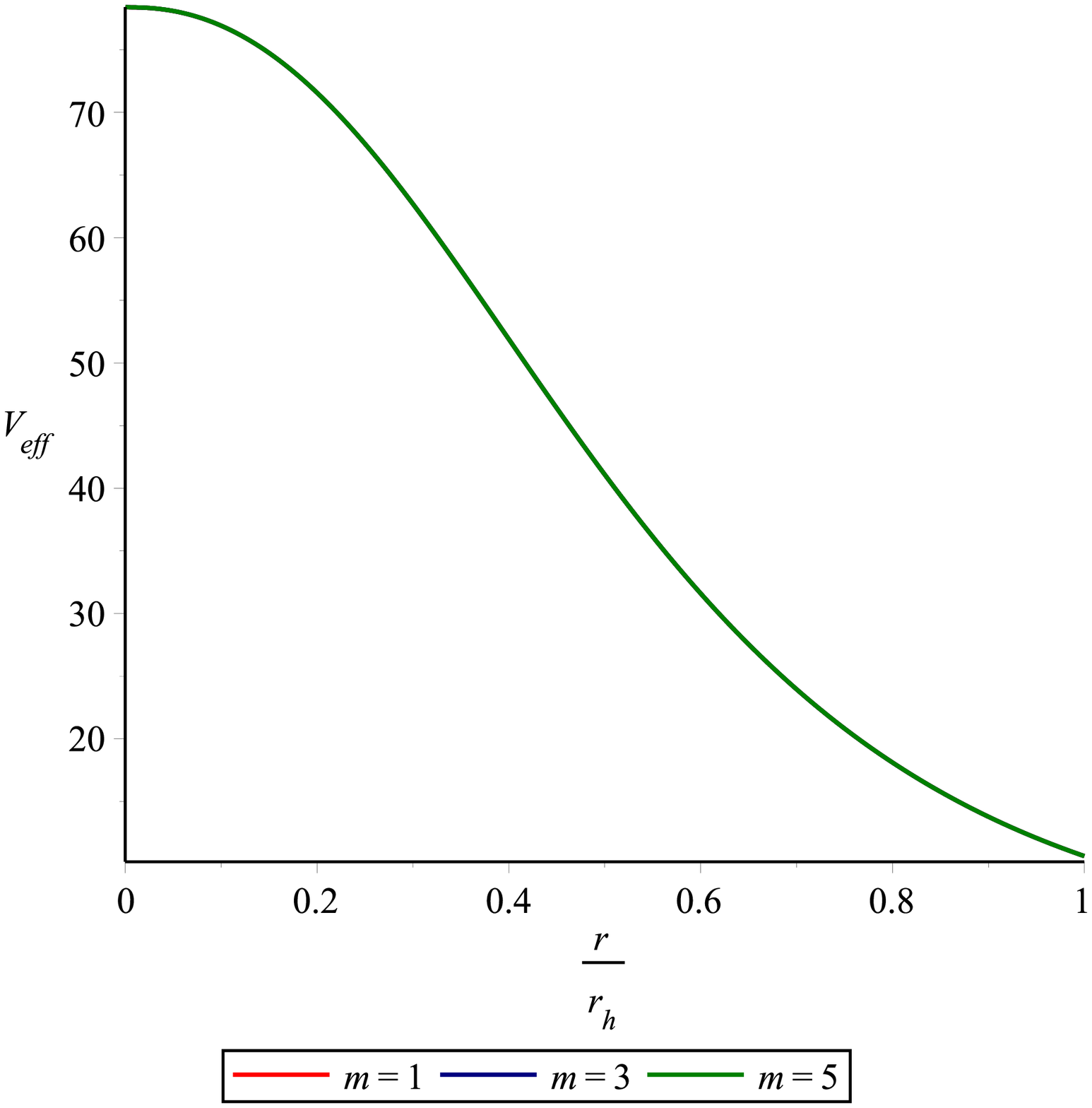,width=0.5\linewidth}
\caption{Effective potential for massless scalar field for $m=1$
(left) and $l=5$ (right) with $w=0.1$, $q=1$, $M=1$, $a=0.6$ and
$\alpha=0.01$.}
\end{figure}
\begin{figure}\center
\epsfig{file=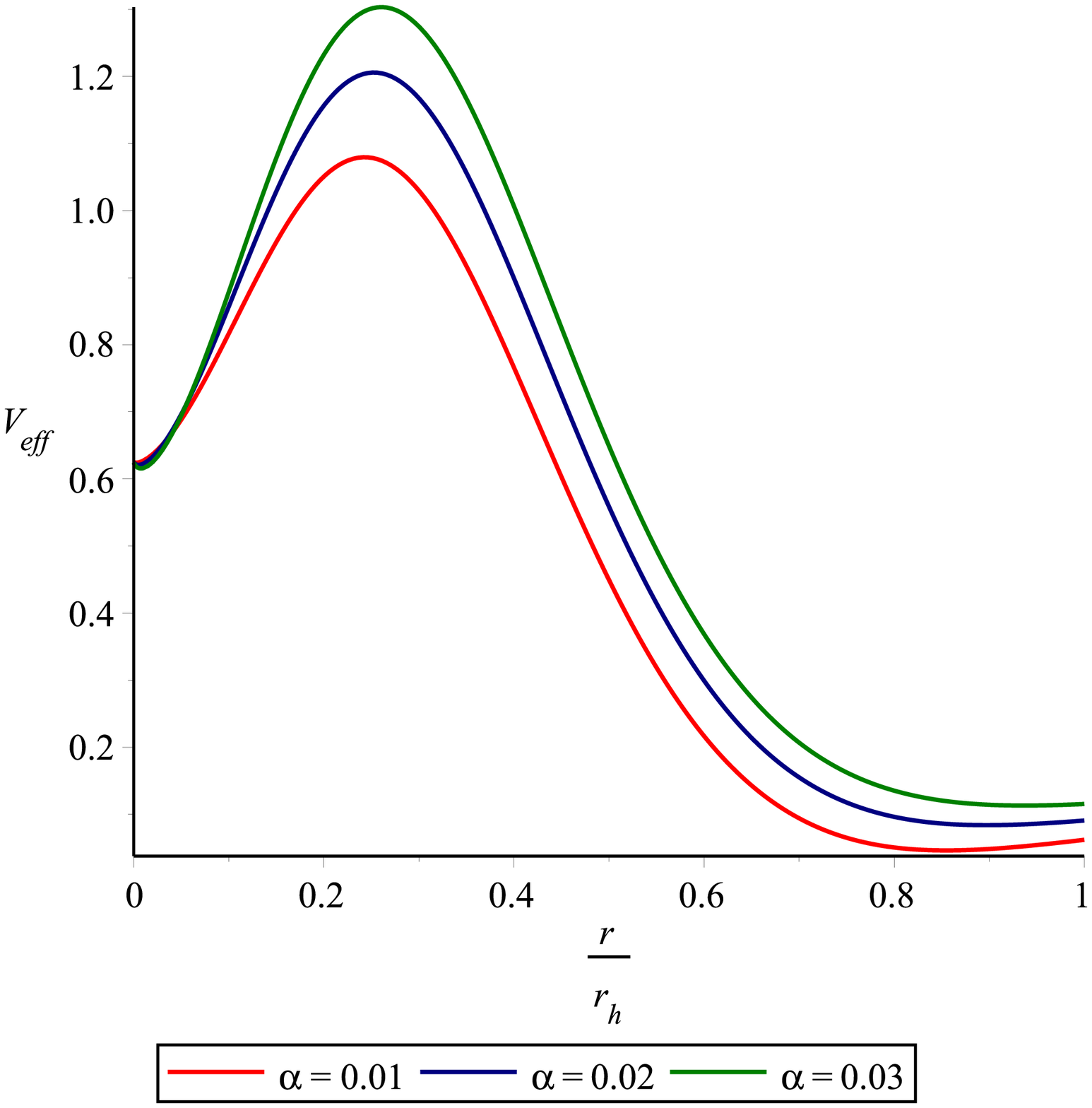,width=0.5\linewidth}\epsfig{file=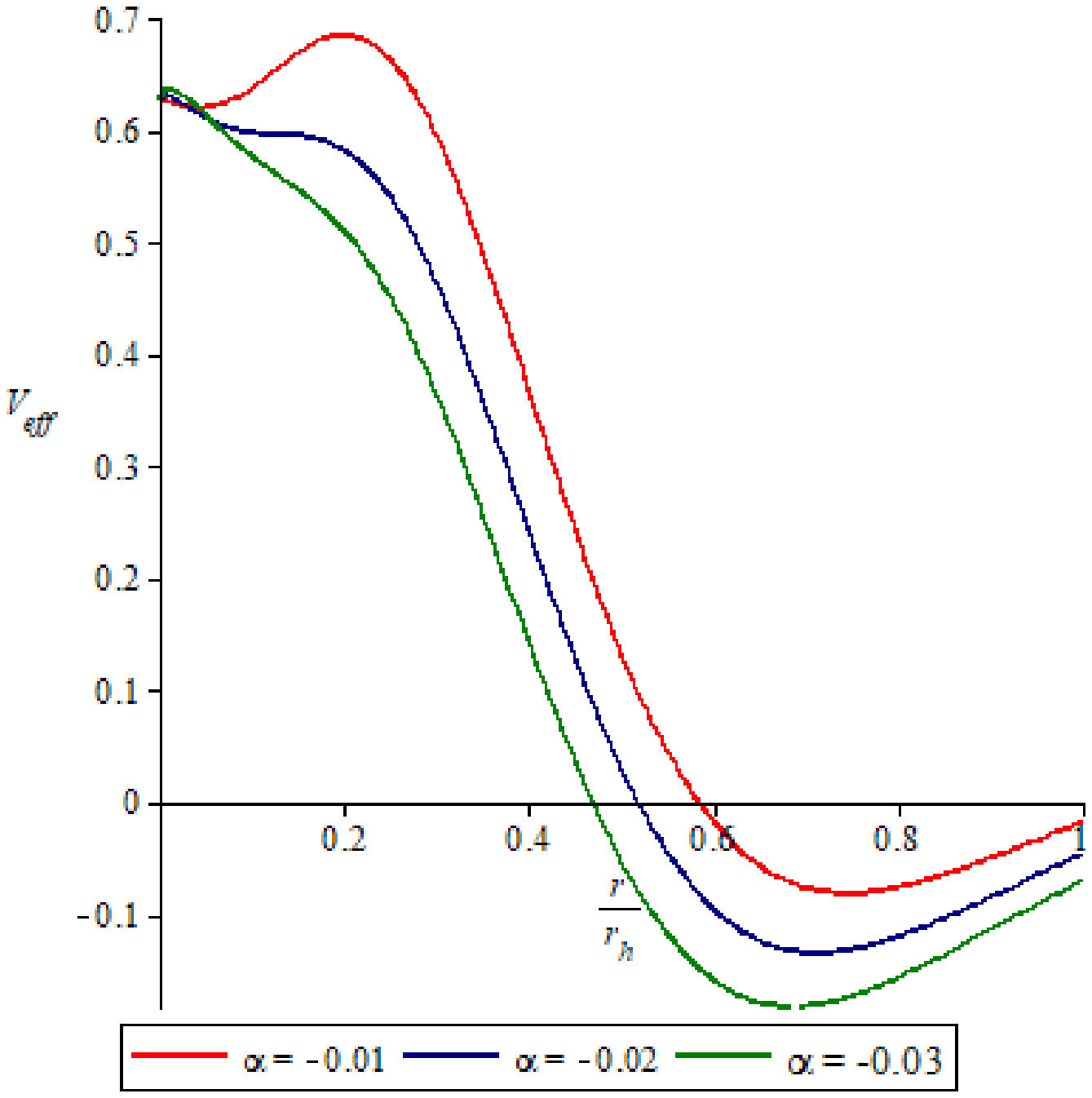,width=0.5\linewidth}
\caption{Effective potential for massless scalar field for
$\alpha>0$ (left) and $\alpha<0$ (right) with $q=1.0$, $a=0.6$,
$m=1$, $M=1$, $l=1$ and $w=0.1$.}
\end{figure}

\section{Greybody Factor}

This section provides the analytic solution of the greybody factor
by using an appropriate technique on the radial equation of motion
(\ref{5}). We determine two asymptotic solutions for different
regimes such as near and far-away from the BH horizon. To obtain the
solution for the whole region, we compare these solutions smoothly
in an intermediate regime.

We choose the following transformation to find the analytic solution
for the near horizon region $r\sim r_{h}$
\begin{eqnarray}\label{6}
r\rightarrow Y=
\frac{a^{2}+r^{2}-\frac{2Mr^{4}}{(r^{2}+q^{2})^{\frac{3}{2}}}+\alpha
r\ln\frac{r}{|\alpha|}}{a^{2}+r^{2}+\alpha r\ln\frac{r}{|\alpha|}},
\end{eqnarray}
which gives
\begin{eqnarray}\nonumber
\frac{d Y}{d r}=\frac{(1-Y)U(r_{h})}{r_{h}(r^{2}_{h}+q^{2})},
\end{eqnarray}
where
\begin{eqnarray}\nonumber
U(r_{h})=\frac{r^{2}_{h}(r^{2}_{h}+\alpha
r_{h}-a^{2})-g^{2}\bigg(2(2a^{2}+r^{2}_{h})-\alpha
r_{h}(1-3\ln\frac{r_{h}}{|\alpha|})\bigg)}{a^{2}+r^{2}_{h}+\alpha
r_{h}\ln\frac{r_{h}}{|\alpha|}}.
\end{eqnarray}
Using these results in the radial equation (\ref{5}), it follows
that
\begin{equation}\label{7}
Y(1-Y)\frac{d^{2}R_{w l m}}{d Y^{2}}+(A-BY)\frac{d R_{w l m}}{d Y}+
\frac{1}{(1-Y)U^{2}}\bigg[\frac{\chi_{h}}{Y}-\lambda_{h}\bigg]R_{w l
m}=0,
\end{equation}
where
\begin{eqnarray}\nonumber
A&=&\frac{(r^{2}_{h}+q^{2})}{r^{2}_{h} U(r_{h})}\frac{d }{d
r}\bigg(r^{3}_{h}Y U(r_{h})\bigg),\quad B=\frac{5
r^{2}}{U(r_{h})},\\\nonumber
\chi_{h}&=&r^{2}_{h}(r^{2}_{h}+q^{2})^{2}\bigg[w^{2}(a^{2}+r^{2}_{h})^{2}+a^{2}
m^{2}-2amw(a^{2}+r^{2}_{h}-\Omega)\bigg],\\\nonumber
\lambda_{h}&=&\frac{r^{2}_{h}(r^{2}_{h}+q^{2})^{2}}{a^{2}+r^{2}_{h}+\alpha
r_{h}\ln\frac{r_{h}}{|\alpha|}}(a^{2} w^{2}+\lambda^{m}_{l}).
\end{eqnarray}

We redefine the function in Eq.(\ref{7}) as
\begin{equation}\nonumber
R_{w l m}(Y)=Y^{\epsilon_{1}}(1-Y)^{\eta_{1}}\hat{F}(Y),
\end{equation}
so that Eq.(\ref{7}) takes the form
\begin{eqnarray}\nonumber
Y(1-Y)\frac{d^{2}\hat{F}(Y)}{d Y^{2}} +
\bigg[2\epsilon_{1}+A-(2\epsilon_{1}+2\eta_{1}+B)Y\bigg]\frac{d
\hat{F}(Y)}{d Y}+ \bigg[\bigg(\epsilon^{2}_{1}-\epsilon_{1}\\+
A\epsilon_{1}+\frac{\chi^{\ast}_{h}}{U^{2}}
\bigg)\frac{1}{Y}+\nonumber
\bigg(\eta^{2}_{1}-\eta_{1}-\eta_{1}A+\eta_{1}B+
\frac{\chi_{h}}{U^{2}}-\frac{\lambda_{h}}{U^{2}}\bigg)\frac{1}
{1-Y}\bigg]\hat{F}(Y)=0.
\end{eqnarray}
The power coefficients $\epsilon_{1}$ and $\eta_{1}$ can be
calculated as
\begin{eqnarray}\nonumber
\epsilon^{2}_{1}-\epsilon_{1}+A\epsilon_{1}+\frac{\chi_{h}}
{U^{2}}=0,\\\nonumber
\eta^{2}_{1}-\eta_{1}-\eta_{1}A+\eta_{1}B+\frac{\chi_{h}}
{U^{2}}-\frac{\lambda_{h}}{U^{2}}=0.
\end{eqnarray}
We finally obtain the hypergeometric (HG) type of differential
equation of Eq.(\ref{5}) as
\begin{equation}\nonumber
Y(1-Y)\frac{d^{2} \hat{F}(Y)}{d
Y^{2}}+\bigg[\bar{c}_{1}-(1+\bar{a}_{1}+\bar{b}_{1})Y\bigg]\frac{d
\hat{F}(Y)}{d Y}-\bar{a}_{1}\bar{b}_{1}\hat{F}(Y)=0,
\end{equation}
where
\begin{eqnarray}\nonumber
\bar{a}_{1}=\eta_{1}+\epsilon_{1}+B-1,\quad\bar{b}_{1}=\eta_{1}+
\epsilon_{1},\quad\bar{c}_{1}= 2\epsilon_{1}+A.
\end{eqnarray}
Its general solution for the near horizon (NH) is given as
\begin{eqnarray}\nonumber
(R_{w l m})_{NH}(Y)&=&
\hat{A_{1}}Y^{\epsilon_{1}}(1-Y)^{\eta_{1}}\hat{F}(\bar{a}_{1},\bar{b}_{1},\bar{c}_{1}
; Y )+\hat{A_{2}}Y^{-\epsilon_{1}}(1-Y)^{\eta_{1}}\\\label{8}
&\times&\hat{F}(1-\bar{c}_{1}+\bar{a}_{1},
1-\bar{c}_{1}+\bar{b}_{1},-\bar{c}_{1}+2;Y),
\end{eqnarray}
where $\hat{A_{1}}$ and $\hat{A_{2}}$ are constants with
\begin{eqnarray}\nonumber
\epsilon^{\pm}_{1}&=& \frac{1}{2}\bigg[(1-A)\pm\sqrt{(1-A)^{2}-4
\frac{\chi_{h}}{U^{2}}}\bigg],\\\nonumber
\eta^{\pm}_{1}&=&\frac{1}{2}\bigg[(1+A-B)\pm\sqrt{(1+A-B)^{2}+4\bigg(
\frac{\lambda_{h}}{U^{2}}-\frac{\chi_{h}}{U^{2}}\bigg)}\bigg].
\end{eqnarray}
Applying the boundary conditions, i.e., no outgoing modes are
observed near the BH horizon. We can choose either $\hat{A_{1}}=0$
or $\hat{A_{2}}=0$ which depends on the choice of $\epsilon_{1}$. It
is found that the constants $\hat{A_{1}}$ and $\hat{A_{2}}$ remain
the same for both values of $\epsilon_{1}$, so that we take
$\epsilon_{1}^{+}=\epsilon_{1}^{-}$ by putting $\hat{A_{2}}=0$. The
signs of $\eta_{1}$ can also be determined similarly by applying the
convergence condition of the HG function which is valid for
$\eta_{1}^{+}=\eta_{1}^{-}$. Thus the analytic solution for the NH
takes the form
\begin{equation}\label{9}
(R_{w l m})_{N
H}(Y)=\hat{A_{1}}Y^{\epsilon_{1}}(1-Y)^{\eta_{1}}\hat{F}(\bar{a}_{1},
\bar{b}_{1},\bar{c}_{1};Y).
\end{equation}

Now we apply the above procedure for NH on the radial equation for
far-away from BH horizon by replacing $\Omega(r)$ with $Z(r)$ as
\begin{equation}\label{10}
Z(r)=1+\frac{a^{2}}{r^{2}}+\frac{\alpha}{r}\ln\frac{r}{|\alpha|}.
\end{equation}
Consequently, the radial equation becomes
\begin{eqnarray}\nonumber
&&Z(1-Z)\frac{d^{2} R_{ w l m}}{dZ^{2}}+(C-D^{\ast}Z)\frac{d  R_{ w
l m}}{d
Z}+\frac{1}{D^{2}(1-Z)}\bigg[\frac{\chi^{\ast}_{h}}{Z}-\lambda^{\ast}_{h}
\bigg] R_{ w l m}=0,
\end{eqnarray}
where
\begin{eqnarray}\nonumber
\chi^{\ast}_{h}&=&\frac{r^{2}Z^{2}}{\Omega^{2}}
\bigg[w^{2}(a^{2}+r^{2})^{2}+a^{2}m^{2}-2amw(a^{2}+r^{2}-\Omega)
\bigg],\\\nonumber \lambda^{\ast}_{h}&=&\frac{Z r^{2}}{\Omega}(a^{2}
w^{2} +\lambda^{m}_{l}),\quad C=\frac{Z[\Omega(1-Z)D]^{'}}{\Omega
Z^{'}D},\quad D^{\ast}=\frac{1}{D}.
\end{eqnarray}
Redefine the field in the above differential equation as
\begin{eqnarray}\nonumber
R_{wlm}(Z)=Z^{\epsilon_{2}}(1-Z)^{\eta_{2}}\hat{F(Z)},
\end{eqnarray}
we have
\begin{eqnarray}\nonumber
&&Z(1-Z)\frac{d^{2}\hat{F(Z)}}{d Z^{2}}+\bigg[ 2\epsilon_{2} + C -
(2\epsilon_{2}+2\eta_{2} + D^{\ast})Z\bigg]\frac{d \hat{F(Z)}}{d Z}
+\bigg[\bigg(\epsilon^{2}_{2}-\epsilon_{2}\\&&+\epsilon_{2}C+
\frac{\chi^{\ast}_{h}}{D^{2}}\bigg)\bigg(
\eta_{2}^{2}-\eta_{2}-\eta_{2}C+\eta_{2} D^{\ast} +
\frac{\chi^{\ast}_{h}}{D^{2}}-\frac{\lambda^{\ast}_{h}}{D^{2}}
\bigg)\frac{1}{1-Z}\bigg]\hat{F(Z)}=0.\nonumber
\end{eqnarray}
The power coefficients $\epsilon_{2}$ and $\eta_{2}$ can be found as
\begin{eqnarray}\nonumber
\epsilon^{2}_{2}-(1-C)\epsilon_{2}+\frac{\chi^{\ast}_{h}}{D^{2}}=0,\\\nonumber
\eta^{2}_{2}-(1+C-D^{\ast})\eta_{2}+\frac{\chi^{\ast}_{h}}
{D^{2}}-\frac{\lambda^{\ast}_{h}}{D^{2}}=0.
\end{eqnarray}
Hence, the above equation in terms of HG form with
\begin{eqnarray}\nonumber
\bar{a}_{2}=\epsilon_{2}+\eta_{2}+D^{\ast}-1,\quad\bar{b}_{2}=
\epsilon_{2}+\eta_{2},\quad\bar{c}_{2}= 2\epsilon_{2}+C,
\end{eqnarray}
becomes
\begin{equation}\nonumber
Z(1-Z)\frac{d^{2}\hat{F(Z)}}{d Z^{2}}+
\bigg[\bar{c}_{2}-(1+\bar{a}_{2}+\bar{b}_{2})\bigg]\frac{d
\hat{F(Z)}}{d Z}-\bar{a}_{2}\bar{b}_{2} \hat{F(Z)}=0.
\end{equation}
Its general solution is
\begin{eqnarray}\nonumber
(R_{wlm})_{f}(Z)&=&
\hat{B_{1}}Z^{\epsilon_{2}}(1-Z)^{\eta_{2}}\hat{F}(\bar{a}_{2},
\bar{b}_{2},\bar{c}_{2};Z)+\hat{B_{2}}Z^{-\epsilon_{2}}
(-Z+1)^{\eta_{2}}\\\label{11}&\times&
\hat{F}(1-\bar{c}_{2}+\bar{a}_{2},1-\bar{c}_{2}+\bar{b}_{2},
-\bar{c}_{2}+2;Z),
\end{eqnarray}
with
\begin{eqnarray}\nonumber
\epsilon_{2}&=& \frac{1}{2}\bigg[(1-C)\pm \sqrt{(1-C)^{2}-
4\frac{\chi^{\ast}_{h}}{D^{2}}}\bigg],\\\nonumber \eta_{2}&=&
\frac{1}{2}\bigg[(1+C-D^{\ast})\pm \sqrt{(1+C-D^{\ast})^{2} -
4\bigg(\frac{\chi^{\ast}_{h}}{D^{2}}
-\frac{\lambda^{\ast}_{h}}{D^{2}}\bigg)}\bigg],
\end{eqnarray}
where $\hat{B_{1}}$ and $\hat{B_{2}}$ are arbitrary constants.
Similar to NH, we choose the HG convergence condition to compare the
solutions with the same choices of
$\epsilon_{2}^{+}=\epsilon_{2}^{-}$ and $\eta_{2}^{+}=\eta_{2}^{-}$.

\section{Matching to an Intermediate Regime}

This section is devoted to matching the obtained solutions (NH and
far-away) efficiently in the intermediate region for all values of
$r$. In this scenario, we expand the solution by changing the HG
function argument $Y$ to $1-Y$ in Eq.(\ref{9}) as
\begin{eqnarray}\nonumber
(R_{w l m})_{N H}(Y)&=& (-Y+1)^{\eta_{1}}
\bigg[\frac{\Gamma(-\bar{a}_{1}-\bar{b}_{1}+\bar{c}_{1})
\Gamma(\bar{c}_{1})}{\Gamma(-\bar{a}_{1}+\bar{c}_{1})
\Gamma(-\bar{b}_{1}+\bar{c}_{1})} \hat{F}(\bar{a}_{1},\bar{b}_{1}
,\bar{c}_{1};1-Y)\\\nonumber&+&
(1-Y)^{-\bar{a}_{1}-\bar{b}_{1}+\bar{c}_{1}}
\frac{\Gamma(\bar{c}_{1})
\Gamma(-\bar{c}_{1}+\bar{b}_{1}+\bar{a}_{1})}{\Gamma(\bar{b}_{1})
\Gamma(\bar{a}_{1})}
\\\nonumber&\times& \hat{F}(+\bar{c}_{1}-\bar{a}_{1},
+\bar{c}_{1}-\bar{b}_{1},1-\bar{a}_{1}-\bar{b}_{1}+\bar{c}_{1} ; 1-Y
)\bigg]\hat{A_{1}}Y^{\epsilon_{1}}.
\end{eqnarray}
Using Eq.(\ref{2}) in (\ref{6}), it follows that
\begin{eqnarray}\nonumber
1-Y = \frac{2Mr^{4}}{(r^{2}+q^{2})^{\frac{3}{2}}(r^{2}+a^{2}+\alpha
r\ln\frac{r}{|\alpha|})}.
\end{eqnarray}
The stretched NH for $Y\rightarrow1$ and limiting value of $r\gg
r_{h}$ yield
\begin{eqnarray}\nonumber
(1-Y)^{\eta_{1}}\simeq \bigg[\frac{r_{h}
(q_{\ast}^{2}+1)^{\frac{3}{2}}(a_{\ast}^{2}+1+\alpha_{\ast}
\ln\frac{r_{h}}{|\alpha|})}{r}\bigg]^{\eta_{1}} \\\nonumber \simeq
\bigg[\frac{r_{h}
(q_{\ast}^{2}+1)^{\frac{3}{2}}(1+a_{\ast}^{2}+\alpha_{\ast}
\ln\frac{r_{h}}{|\alpha|})}{r}\bigg]^{-l},
\end{eqnarray}
and
\begin{eqnarray}\nonumber
(1-Y)^{\eta_{1}-\bar{a}_{1}-\bar{b}_{1}+\bar{c}_{1}}&\simeq &
\bigg[\frac{r_{h}
(1_{\ast}^{2}+1)^{\frac{3}{2}}(1+a_{\ast}^{2}+\alpha_{\ast}
\ln\frac{r_{h}}{|\alpha|})}{r}\bigg]^{-\eta_{1}+A-B+1},\\\nonumber
&\simeq & \bigg[\frac{r_{h}
(q_{\ast}^{2}+1)^{\frac{3}{2}}(1+a_{\ast}^{2}+\alpha_{\ast}
\ln\frac{r_{h}}{|\alpha|})}{r}\bigg]^{1+l},
\end{eqnarray}
where $a_{\ast}=\frac{a}{r}$, $q_{\ast}=\frac{q}{r}$ and
$\alpha_{\ast}=\frac{\alpha}{r}$. We would like to mention here that
all the above constraints are valid for smaller values of charge and
rotation parameter. In an intermediate zone, the NH solution can be
expressed as
\begin{equation}\label{12}
(R_{w l m})_{NH}(Y)=\tilde{A_{1}}\bigg(\frac{r}{r_{h}}
\bigg)^{-l}+\tilde{A_{2}}\bigg(\frac{r}{r_{h}}\bigg)^{-(1+l)},
\end{equation}
with
\begin{eqnarray}\nonumber
\tilde{A_{1}}&=&\hat{A_{1}}\bigg[(1+q_{\ast}^{2})^{\frac{3}{2}}
(1+a_{\ast}^{2}+\alpha_{\ast}\ln\frac{r_{h}}{|\alpha|}) \bigg]^{-l}
\frac{\Gamma(\bar{c}_{1})\Gamma(-\bar{a}_{1}-\bar{b}_{1}+\bar{c}_{1})
}{\Gamma(+\bar{c}_{1}-\bar{b}_{1})\Gamma(+\bar{c}_{1}-
\bar{a}_{1})},\\\nonumber \tilde{A_{2}}&=&
\bigg[(q_{\ast}^{2}+1)^{\frac{3}{2}}(1+a_{\ast}^{2}+\alpha_{\ast}
\ln\frac{r_{h}}{|\alpha|})\bigg]^{1+l}
\frac{\Gamma(\bar{a}_{1}+\bar{b}_{1}-\bar{c}_{1})
\Gamma(\bar{c}_{1})}{\Gamma(\bar{a}_{1})\Gamma(\bar{b}_{1})}.
\end{eqnarray}

Now, we find the solution far away from the BH event horizon and
stretch the HG function arguments by changing $Z$ with $1-Z$ with
$Z(r_{f})\rightarrow 0$. Hence, Eq.(\ref{11}) reduces to
\begin{eqnarray}\nonumber
(1-Z)^{\eta_{2}}\simeq \bigg(\frac{\ln\frac{r}{|\alpha|}}
{\ln\frac{r_{f}}{|\alpha|}}\bigg)^{-l} \bigg(\frac{r}{r_{f}}
\bigg)^{l},
\end{eqnarray}
and
\begin{eqnarray}\nonumber
(1-Z)^{\eta_{2}+\bar{c}_{2}-\bar{a}_{2}-\bar{b}_{2}}\simeq\bigg(
\frac{r}{r_{f}}\bigg)^{-(1+l)}\bigg(
\frac{\ln\frac{r}{|\alpha|}}{\ln\frac{r_{f}}{|\alpha|}}\bigg)^{1+l}.
\end{eqnarray}
We restrict the parameters $a$ and $q$ to smaller values for the
far-field horizon and the solution of Eq.(\ref{11}) yields
\begin{equation}\label{13}
(R_{w l m})_{f}(Z)=\bigg( \tilde{H_{1}}\tilde{B_{1}}+ \tilde{H_{2}}
\tilde{B_{2}}\bigg)\bigg(\frac{r}{r_{f}}\bigg)^{l}+\bigg(
\tilde{H_{3}}\tilde{B_{1}}+ \tilde{H_{4}}
\tilde{B_{2}}\bigg)\bigg(\frac{r}{r_{f}}\bigg)^{-(1+l)},
\end{equation}
where
\begin{eqnarray}\nonumber
\tilde{H_{1}}&=&\frac{\Gamma(\bar{c}_{2})
\Gamma(\bar{c}_{2}-\bar{a}_{2}-\bar{b}_{2})}
{\Gamma(\bar{c}_{2}-\bar{a}_{2})\Gamma(\bar{c}_{2}-\bar{b}_{2})}
\bigg(\frac{\ln\frac{r}{|\alpha|}}
{\ln\frac{r_{f}}{|\alpha|}}\bigg)^{-l},\\\nonumber\tilde{H_{2}} &=&
\frac{\Gamma(2-\bar{c}_{2})
\Gamma(-\bar{a}_{2}-\bar{b}_{2}+\bar{c}_{2})}{\Gamma(1-\bar{a}_{2})
\Gamma(1-\bar{b}_{2})} \bigg(\frac{\ln\frac{r}{|\alpha|}}
{\ln\frac{r_{f}}{|\alpha|}}\bigg)^{-l},\\\nonumber\tilde{H_{3}}&=&
\frac{
\Gamma(\bar{c}_{2})\Gamma(-\bar{c}_{2}+\bar{b}_{2}+\bar{a}_{2}) } {
\Gamma(\bar{b}_{2})\Gamma(\bar{a}_{2})}\bigg(
\frac{\ln\frac{r}{|\alpha|}}
{\ln\frac{r_{f}}{|\alpha|}}\bigg)^{1+l},\\\nonumber\tilde{H_{4}} &=&
\frac{\Gamma(2-\bar{c}_{2})
\Gamma(-\bar{c}_{2}+\bar{a}_{2}+\bar{b}_{2})}
{\Gamma(\bar{a}_{2}-\bar{c}_{2}+1)
\Gamma(\bar{b}_{2}-\bar{c}_{2}+1)}\bigg(
\frac{\ln\frac{r}{|\alpha|}}
{\ln\frac{r_{f}}{|\alpha|}}\bigg)^{1+l}.
\end{eqnarray}

We compare both the asymptotic solutions with the similar powers of
$l$ and $1+l$ as
\begin{eqnarray}\nonumber
\tilde{A_{1}}=\tilde{H_{1}}\tilde{B_{1}}+\tilde{H_{2}}
\tilde{B_{2}},\quad\tilde{A_{2}}=\tilde{H_{3}}\tilde{B_{1}} +
\tilde{H_{4}}\tilde{B_{2}} .
\end{eqnarray}
The integration constants $\tilde{B_{1}}$ and $\tilde{B_{2}}$ are
found to be
\begin{eqnarray}\nonumber
\tilde{B_{2}}=\frac{\tilde{A_{1}}\tilde{H_{3}}-\tilde{A_{2}}
\tilde{H_{1}}}{\tilde{H_{2}}\tilde{H_{3}}-\tilde{H_{1}}
\tilde{H_{4}}},\quad \tilde{B_{1}}=\frac{\tilde{A_{1}}
\tilde{H_{4}}-\tilde{A_{2}}\tilde{H_{2}}}{\tilde{H_{1}}
\tilde{H_{4}}-\tilde{H_{2}}\tilde{H_{3}}}.
\end{eqnarray}
In order to calculate the emission rate of massless scalar field, we
have the expression of greybody factor as \cite{39}
\begin{eqnarray}\nonumber
|A_{l,m}|^{2} =
1-\bigg|\frac{\tilde{B_{2}}}{\tilde{B_{1}}}\bigg|^{2}=
1-\bigg|\frac{\tilde{A_{2}}\tilde{H_{1}}-\tilde{A_{1}}
\tilde{H_{3}}}{\tilde{A_{1}}\tilde{H_{4}}-\tilde{A_{2}}
\tilde{H_{2}}}\bigg|^{2}.
\end{eqnarray}
This is a combination of greybody factor (absorption probability)
and emission rate of massless scalar field derived from the rotating
regular Bardeen BH. It is noted that the waves passing through
far-away from the horizon will face the barrier which works as a
filter to either move forward or reflect. It is a relative relation
between frequency and effective potential. The frequency of the wave
must be larger than the effective potential to cross the barrier. If
the potential exceeds the wave frequency, some portions are
reflected while some may cross the barrier and consequently, the
greybody factor displays a negative trend.

We sketch the above expression for different parameters to discuss
viability of the greybody factor numerically. The effects of
parameters $a$ and $l$ are analyzed on the profile of the greybody
factor. Figure \textbf{4} indicates that the absorption rate of
scalar field increases for higher values of the rotation parameter
as well as angular momentum. Figure \textbf{5} shows the influence
of $q$ for the absorption rate of the massless scalar field. It is
found that BH absorbs partial waves with the increasing values of
magnetic charge which increases the greybody factor and reduces the
emission rate of the scalar field. The impact of PFDM parameter
$\alpha$ on the greybody factor is shown in Figure \textbf{6}. It is
noted that increasing the value of $\alpha$ yields a high rate of
absorption probability of initial waves and raises the greybody
factor.
\begin{figure}
\epsfig{file=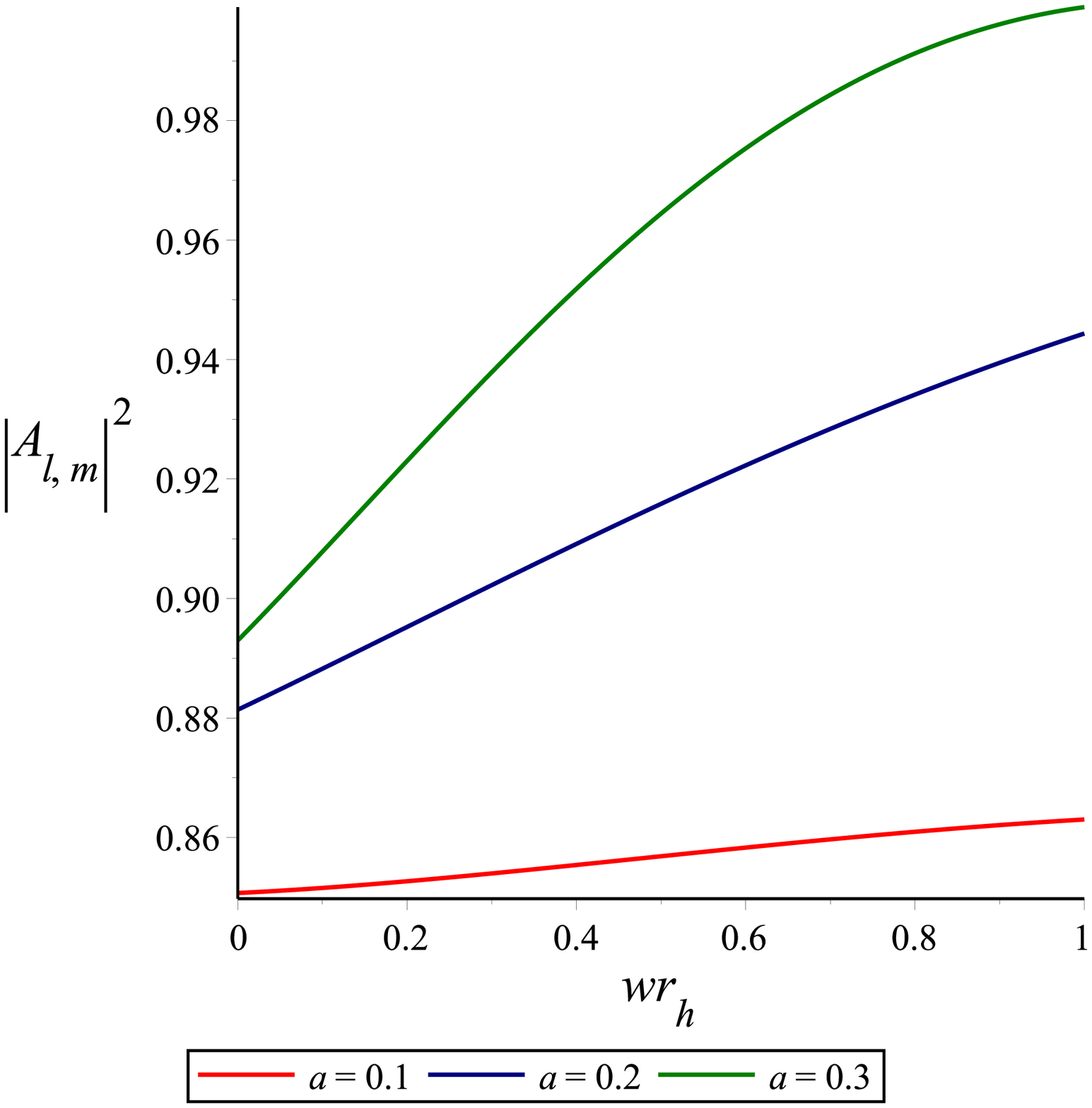,width=.5\linewidth}
\epsfig{file=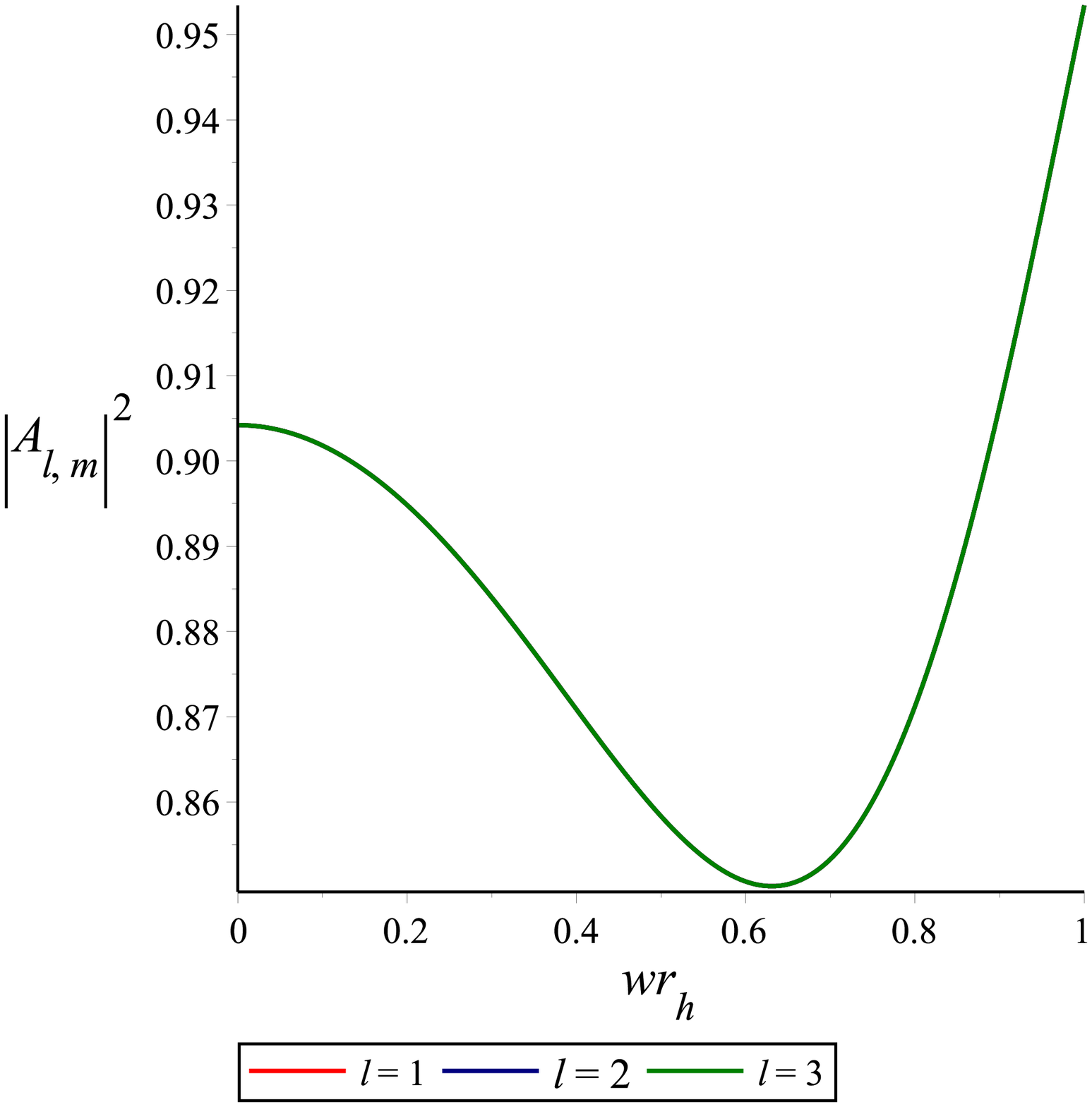,width=.5\linewidth}\caption{Greybody factor for
massless scalar field for $l=1$ (left) and $a=0.1$ (right) with
$r=0.5$, $m=1$, $M=1$, $q=0.1$ and $\alpha=0.01$.}
\end{figure}
\begin{figure}
\epsfig{file=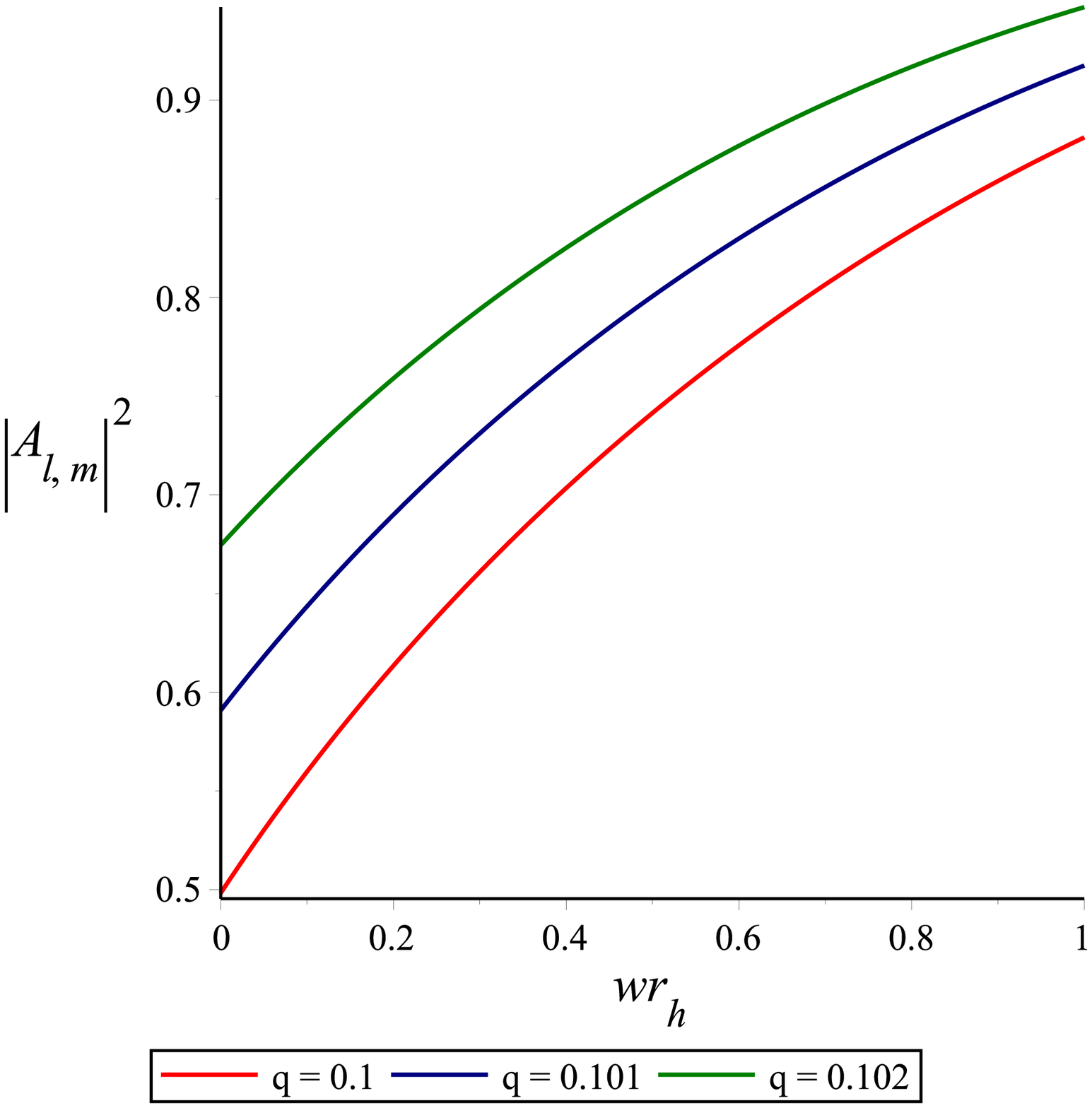,width=.5\linewidth}
\epsfig{file=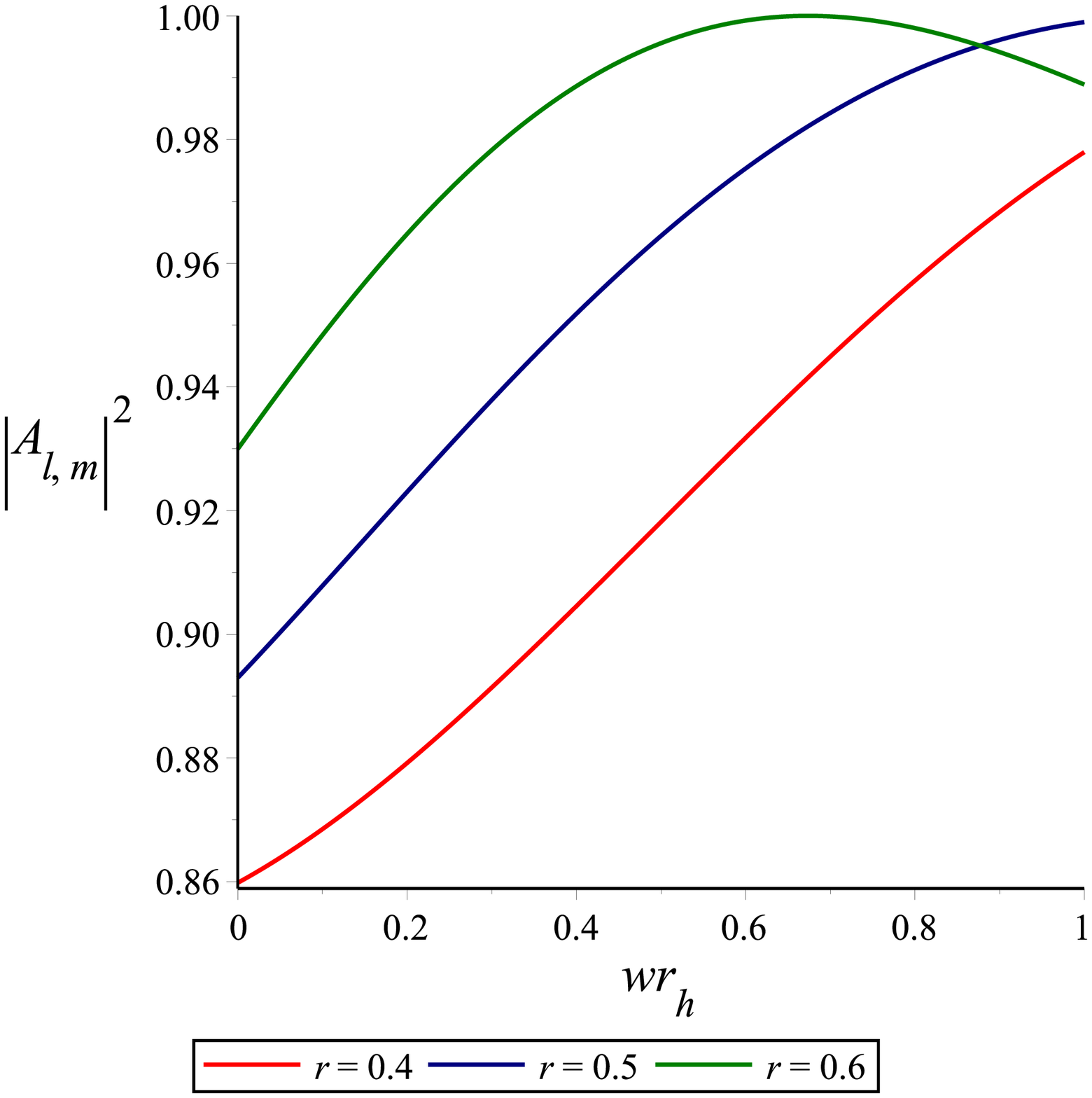,width=.5\linewidth}\caption{Greybody factor for
massless scalar field for $r=0.4$ (left) and $q=0.1$ (right) with
$a=0.1$, $l=1$, $m=1$, $M=1$, $\alpha=0.01$.}
\end{figure}
\begin{figure}
\epsfig{file=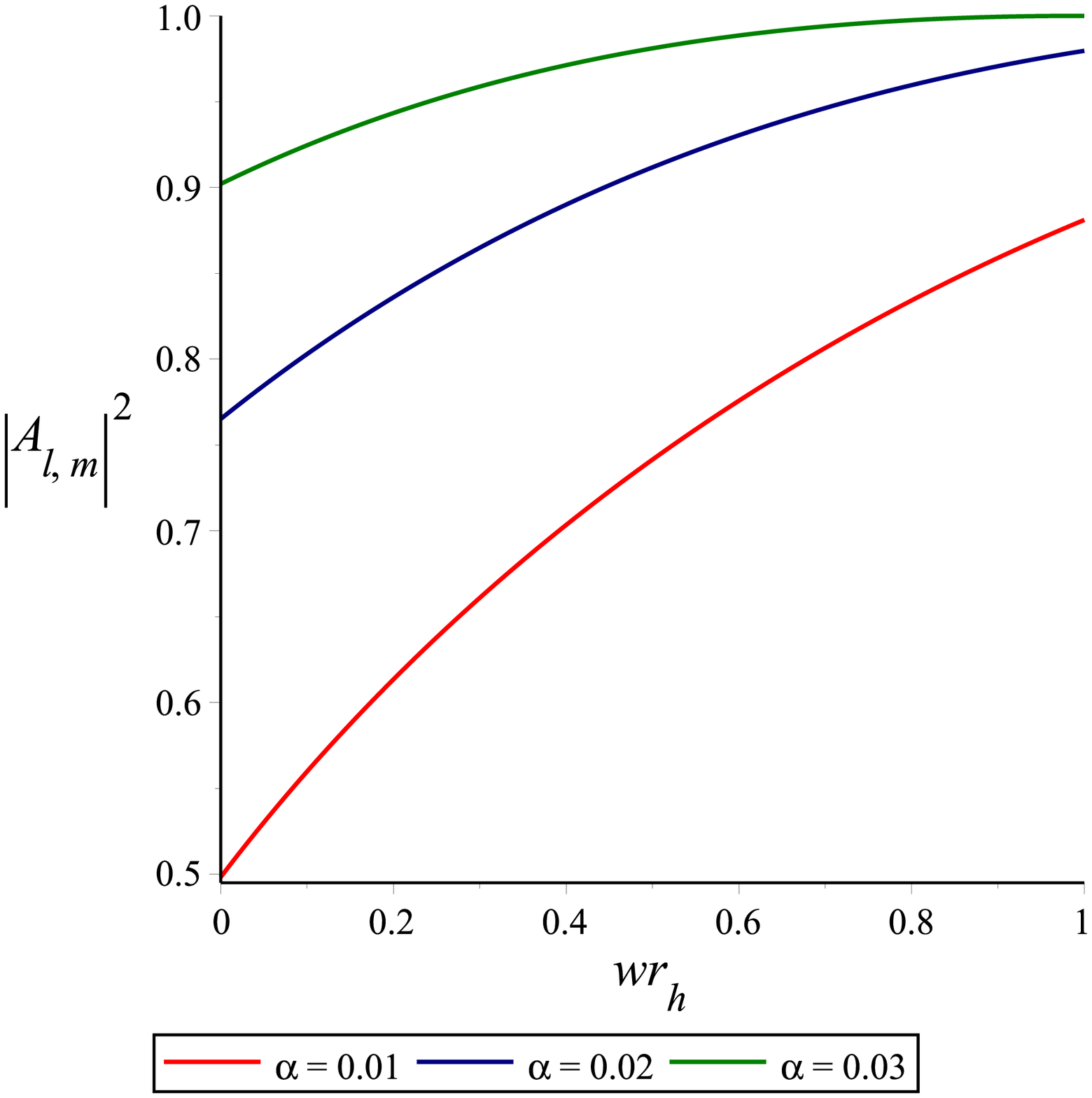,width=.5\linewidth}
\epsfig{file=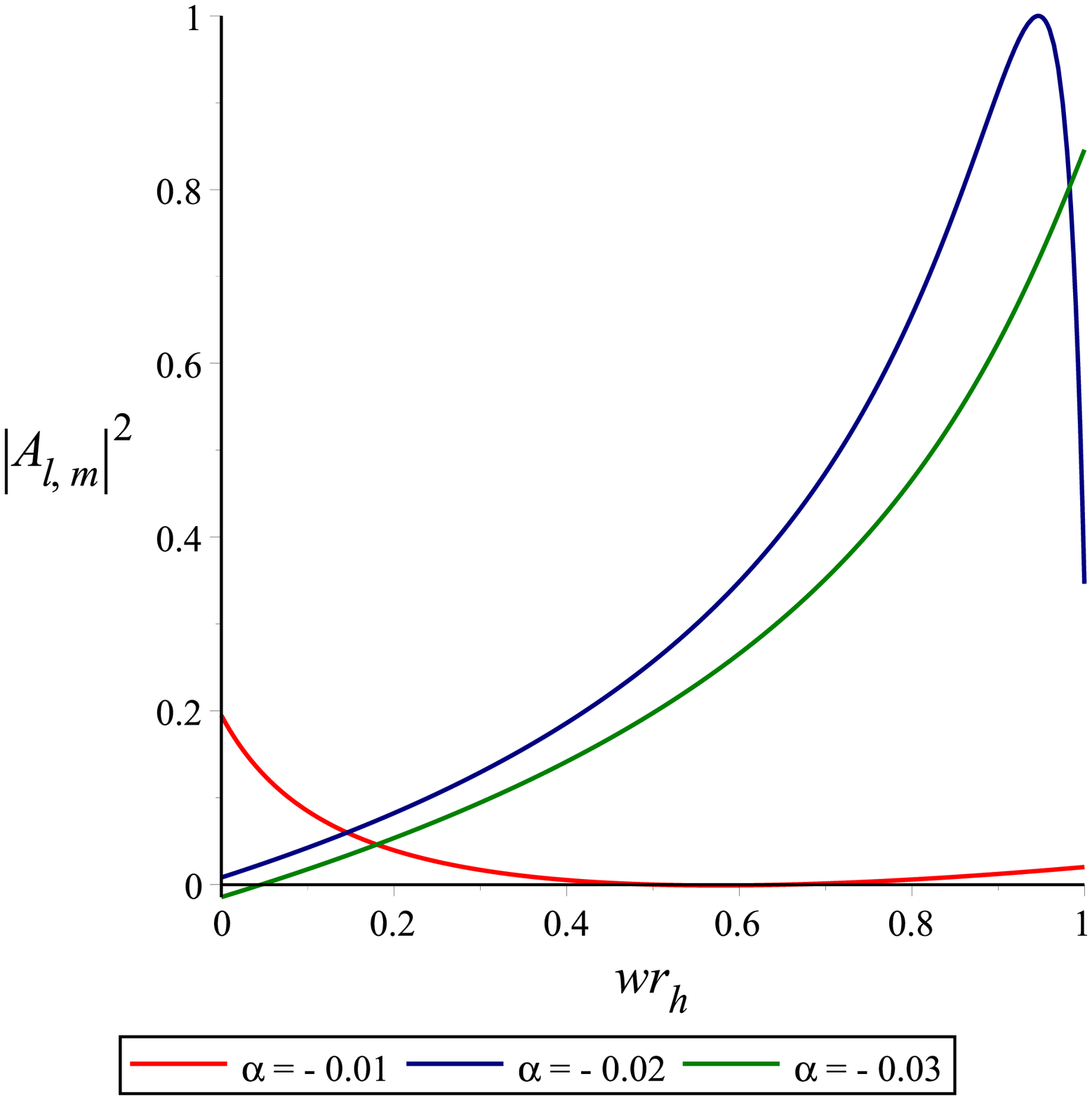,width=.5\linewidth}\caption{Greybody factor
for massless scalar field for $\alpha>0$ (left) and $\alpha<0$
(right) with $a=0.1$, $r=0.1$, $l=1$, $m=1$, $M=1$ and $q=0.1$.}
\end{figure}

The total amount of massless scalar particles discharged per unit
time and frequency from a BH (particle flux) can be found as
\begin{eqnarray}\nonumber
\frac{d^{2}\widetilde{P}}{dt dw}=\sum_{l, m}|A_{l,m}|^{2}
\frac{1}{2\pi e^{-1+\frac{\kappa}{T_{H}}}}=\sum_{l, m}
\bigg(1-\bigg|\frac{\tilde{A_{1}}\tilde{H_{4}}-\tilde{A_{2}}
\tilde{H_{2}}}{\tilde{A_{1}}\tilde{H_{3}}-\tilde{A_{2}}
\tilde{H_{1}}}\bigg|^{2}\bigg)\frac{1}{2\pi
e^{-1+\frac{\kappa}{T_{H}}}},
\end{eqnarray}
and
\begin{eqnarray}\nonumber
\frac{d^{2} \widetilde{N}}{dt dw}=\sum_{l, m}|A_{l,m}|^{2}
\frac{w}{2\pi e^{-1+\frac{\kappa}{T_{H}}}}=\sum_{l, m}
\bigg(1-\bigg|\frac{\tilde{A_{1}}\tilde{H_{4}}-\tilde{A_{2}}
\tilde{H_{2}}}{\tilde{A_{1}} \tilde{H_{3}}-\tilde{A_{2}}
\tilde{H_{1}}}\bigg|^{2}\bigg)\frac{w}{2\pi
e^{-1+\frac{\kappa}{T_{H}}}}.
\end{eqnarray}
Also, we can have
\begin{eqnarray}\nonumber
\kappa= -\frac{am}{a^{2}+r^{2}}+w, \quad T_{H}=
\frac{1-\frac{a^{2}}{r^{2}}}{4\pi r_{h}(1+\frac{a^{2}}{r^{2}})}.
\end{eqnarray}
The differential equation for the emission rate of angular momentum
can also be written in a similar way. The absorption cross-section
of each partial wave can be calculated as
\begin{eqnarray}\nonumber
\sigma= \sum_{l, m}\frac{\pi}{w^{2}} |A_{l,m}|^{2}=\sum_{l,
m}\frac{\pi}{w^{2}} \bigg(1-\bigg|\frac{\tilde{A_{1}}\tilde{H_{4}}-
\tilde{A_{2}}\tilde{H_{2}}}{\tilde{A_{1}}\tilde{H_{3}}
-\tilde{A_{2}}\tilde{H_{1}}}\bigg|^{2}\bigg).
\end{eqnarray}

\section{Concluding Remarks}

In this paper, we have formulated the analytic model of the greybody
factor for rotating Bardeen BH surrounded by PFDM. We have first
calculated the effective potential using the Klein-Gordon equation
and examined the graphical analysis for different parameters. The
Klein-Gordon equation gives spheroidal and radial equations. The
radial equation is then used to determine two asymptotic solutions
near and far-away from to BH horizon. These two solutions are
matched smoothly in an intermediate regime to find the general
expression of the greybody factor in low rotation and low energy
regimes. We have also evaluated the absorption cross-section and
emission rate for a massless scalar field.

We have analyzed the effective potential and greybody factor for
different values of physical parameters in a low energy regime. It
is found that the barrier's height and absorption rate increase with
large values of the rotation parameter. This reduces the emission
rate of radiations and raises its probability of absorption. We have
seen that the greybody factor increases with the effects of angular
momentum as well as magnetic charge. We have found that BH does not
emit radiations through the barrier but absorb rapidly which raises
the greybody factor. It is found that higher modes of PFDM parameter
yields a high rate of absorption probability. We note that rotating
and non-rotating regular BHs evaporate rapidly as compared to other
BHs as they emit thermal flux of quantum level particles. Hence the
rotating Bardeen BH surrounded by PFDM would not squeeze and
disappear faster as it has the ability to absorb massless scalar
field particles.

It is interesting to mention here that the absorption rate of scalar
field remains the same for both rotating Bardeen as well as Kerr BHs
\cite{40}. It is known that PFDM parameter for non-rotating Bardeen
BH increases the emission rate but decreases the absorption rate of
scalar field \cite{35}. However, we have found the opposite behavior
for the rotating case where the PFDM parameter decreases the
emission rate but increases the absorption rate of scalar field. We
conclude that the the PFDM parameter of rotating Bardeen BH with
PFDM increases the greybody factor. When the rotation parameter
vanishes, our results reduce to the results given in \cite{35}. It
is worthwhile to mention here that all our results reduce to the
corresponding results of Schwarzschild BH when $a=0=\alpha=q$.

\end{document}